\documentclass[english]{spie}
\usepackage[T1]{fontenc}
\usepackage[latin9]{inputenc}
\usepackage{verbatim}
\usepackage{graphicx}

\makeatletter

\providecommand{\tabularnewline}{\\}

\pdfoutput=1

\makeatother

\usepackage{babel}

\begin{document}

\title{Corrugated Silicon Platelet Feed Horn Array for CMB Polarimetry at
150~GHz}

\author{Joseph W. Britton\supit{a,b}, John P. Nibarger\supit{b}, Ki Won
Yoon\supit{b}, James A. Beall\supit{b}, Dan Becker\supit{b}, Hsiao-Mei
Cho\supit{b}, Gene C. Hilton\supit{b}, Johannes Hubmayr\supit{b},
Michael D. Niemack\supit{b}, and Kent D. Irwin\supit{b} \skiplinehalf\supit{a}NIST, Time and Frequency Division;\skiplinehalf\supit{b}NIST, Quantum Electrical Metrology Division}

\authorinfo{National Institute of Standards and Technology (NIST), 325 Broadway,
Boulder, CO, USA -- Corresponding author: J. W. B.: E-mail: britton@nist.gov}

\maketitle
%
{}

Next generation cosmic microwave background (CMB) polarization anisotropy
measurements will feature focal plane arrays with more than $600$
millimeter-wave detectors. We make use of high-resolution photolithography
and wafer-scale etch tools to build planar arrays of corrugated platelet
feeds in silicon with highly symmetric beams, low cross-polarization
and low side lobes. A compact Au-plated corrugated Si feed designed
for $150$~GHz operation exhibited performance equivalent to that
of electroformed feeds: $\sim-0.2$~dB insertion loss, $<-20$~dB
return loss from $120$~GHz to $170$~GHz, $<-25$~dB side lobes
and $<-23$~dB cross-polarization. We are currently fabricating a
$50\,\mbox{mm}$ diameter array with $84$ horns consisting of $33$
Si platelets as a prototype for the SPTpol and ACTpol telescopes.
Our fabrication facilities permit arrays up to $150\,\mbox{mm}$ in
diameter.


\section{INTRODUCTION}

\begin{figure}[b]
\begin{centering}
\includegraphics[width=0.5\columnwidth]{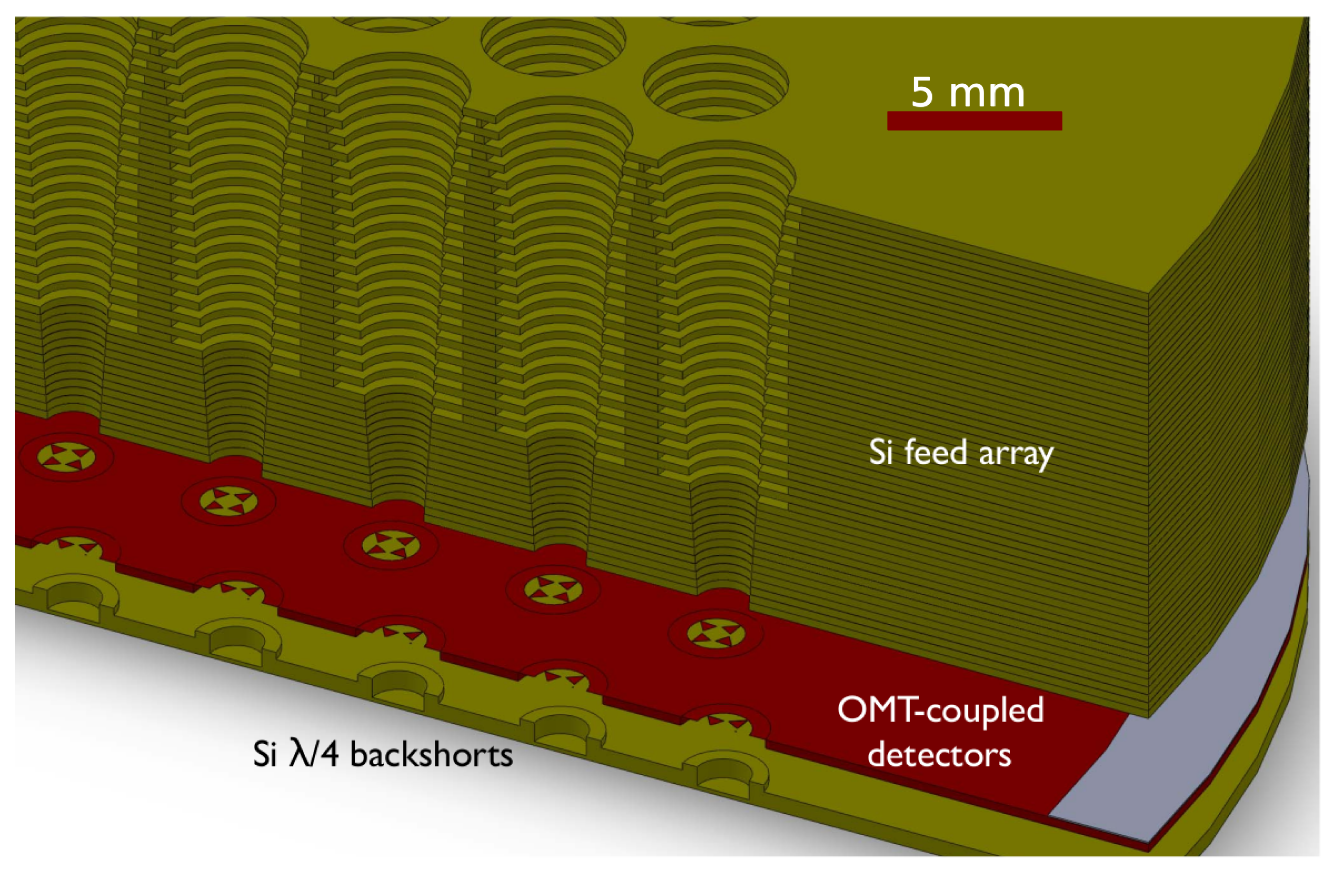}\caption{Illustration showing coupling between a monolithic corrugated Si platelet
feed array and a monolithic array of superconducting detectors fabricated
on Si.\cite{yoon2009ltd13proceedings} The detectors and Si feed array
are in direct thermal contact; there is no thermal expansion mismatch.\cite{britton2009ltd13siHornArray} }

\par\end{centering}

\label{fig:hornArrayAndDetectorArray}
\end{figure}

At present, all published measurements of the CMB polarization used
discrete corrugated feeds to couple free-space to detectors.\cite{wollack2009opticalCouplingForCMB,padin2001cmbCBI,kovak2002cmbpolDASI,barnes2002cmbWMAPhorns,jones2003cmbBoomerang,barkats2005cmbpolCAPMAP,hinderks2009quad}
The use of corrugated feeds is motivated by the need for wide bandwidth,
good beam symmetry, minimal side lobes and low cross polarizations
while maintaining excellent transmission efficiency.\cite{clarricoats1984corrugatedHorns}
Next-generation imaging CMB polarimeters will use monolithic focal
plane detector arrays with hundreds of tightly packed detectors.\cite{lee2008polarbearCMBpol,niemack2010ACTpol,orlando2010BicepAndSpiderTESpolArrays,yoon2010spieTesPolarimetersForCMB}.
In this paper we describe one approach for producing monolithic arrays
of densely packed corrugated scalar feeds using micromachined Au-plated
Si which can be directly contacted to a monolithic array of superconducting
detectors (see Figure~\ref{fig:hornArrayAndDetectorArray}). 

Extension of the conventional electroform approach to fabrication
of monolithic feed arrays is impractical. An alternative is metal
platelet arrays which consist of stacks of perforated metal plates
whose apertures' geometries define the horns' cross sections. A metal
platelet array with adequate performance for CMB polarimetry was demonstrated
with $91$-pixels at $95\,\mbox{GHz}$.\cite{bock2009opticalCouplingForCMB} 

At NIST we are pursuing a new approach to fabricating monolithic corrugated
platelet arrays.\cite{britton2009ltd13siHornArray} Each layer in
the array is a Si wafer with photolithographically defined apertures.
Once assembled and Au-plated, these horn arrays are expected to feature
the same benefits as metal platelet arrays (including high thermal
conductivity) with the following additional advantages: (a) a thermal
expansion match to Si detector arrays, (b) more precise geometry reproduction
(and greater packing density), (c) smaller gaps between platelets
and (d) a straightforward path to arrays of thousands of feeds.\cite{britton2009ltd13siHornArray}
See Table~\ref{table:materialProperties} for a comparison of Si
with metals commonly used in platelet arrays. 

To demonstrate the feasibility of our approach, in 2009 we fabricated
and tested two corrugated feeds made of Si.\cite{britton2009ltd13siHornArray}
Subsequently a prototype monolithic Si array of $73$~corrugated
waveguides was also fabricated and tested. Currently underway at NIST
is the fabrication of a $50\,\mbox{mm}$ diameter array with $84$~horns
consisting of $33$~Si platelets. The platelets composing these devices
were machined in $76.2$~mm and $100\,\mbox{mm}$ diameter Si wafers
by use of photolithography and deep reactive ion etching (DRIE, see
Appendix.~A). 

\begin{table}
\begin{centering}
\begin{tabular}{c|ccccc}
 & $\rho\,\,[\mbox{kg}\cdot\mbox{m}^{-3}]$ & $c_{p}\,\,[\mbox{J}\cdot\mbox{m}^{-3}\cdot\mbox{K}^{-1}/10^{-6}]$  & \multicolumn{2}{c}{$\alpha\,\,[K^{-1}/10^{-6}]$ } & \multicolumn{1}{c}{$\lambda\,\,[W\cdot m^{-1}\cdot K^{-1}]$}\tabularnewline
 & $\sim$300~K & $\sim$300 K & 100 K & 293 K & 297 K\tabularnewline
\hline 
Al  & 2698 & 2.37 & 12.2 & 23.1 & 236\tabularnewline
Cu  & 8933 & 3.39 & 10.3 & 16.5 & 403\tabularnewline
Si  & 2329 & 1.58 & -0.4 & 2.6 & 168 (at 173 K)\tabularnewline
\end{tabular}
\par\end{centering}

\caption{Comparison of structural materials used for horn arrays: $\rho$ is
density\cite{kayelaby1995constantsWeb}, $c_{p}$ is specific heat
(by volume)\cite{kayelaby1995constantsWeb}, $\alpha$ is coefficient
of thermal expansion\cite{kayelaby1995constantsWeb}, and $\lambda$
is thermal conductivity \cite{kayelaby1995constantsWeb}. }

\begin{centering}
\label{table:materialProperties} 
\par\end{centering}

\end{table}

\section{CORRUGATED FEED DESIGN}

\label{sec:fhDesign}

\begin{figure}[b]
\begin{centering}
\includegraphics[width=0.6\columnwidth]{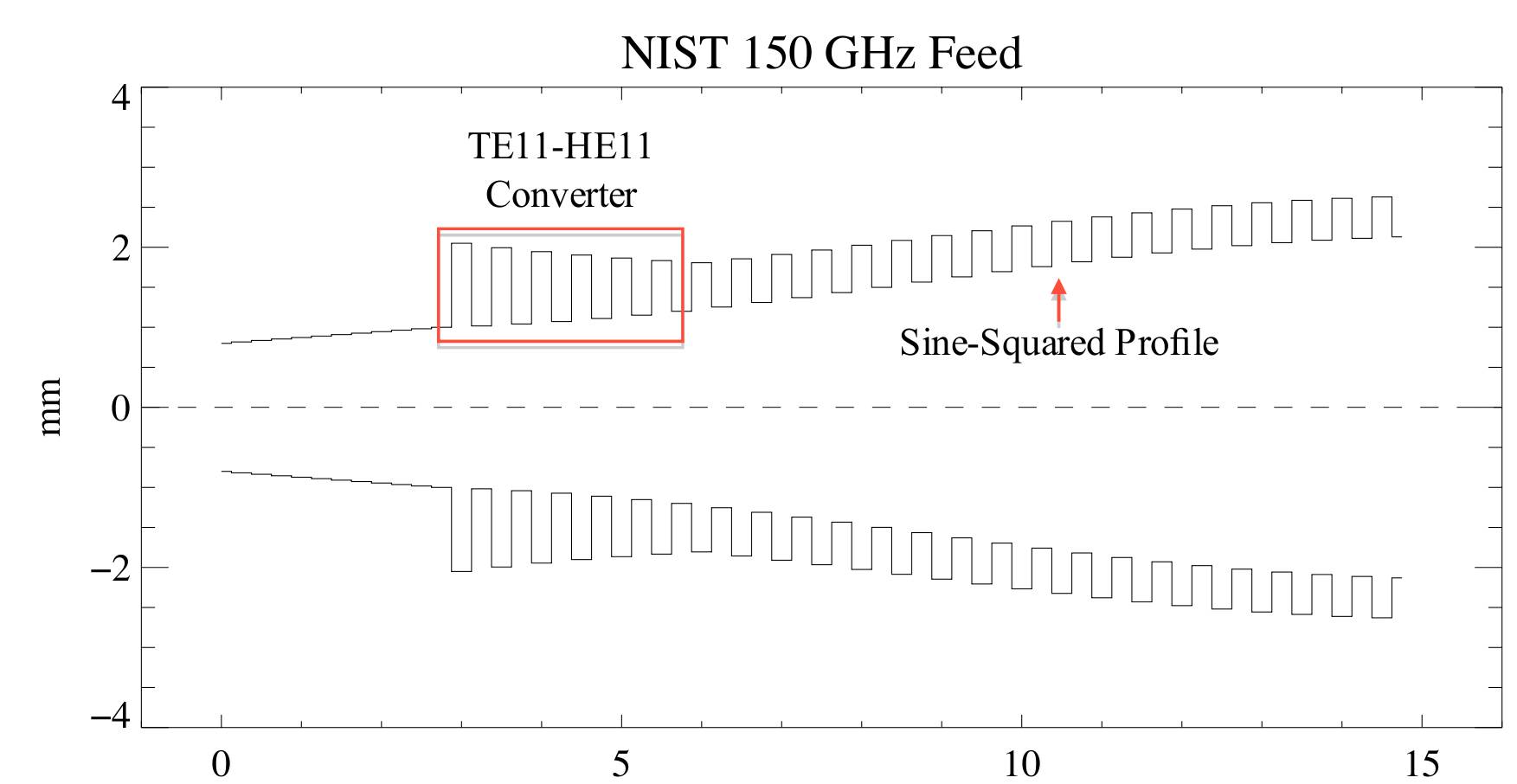}\includegraphics[width=0.4\columnwidth]{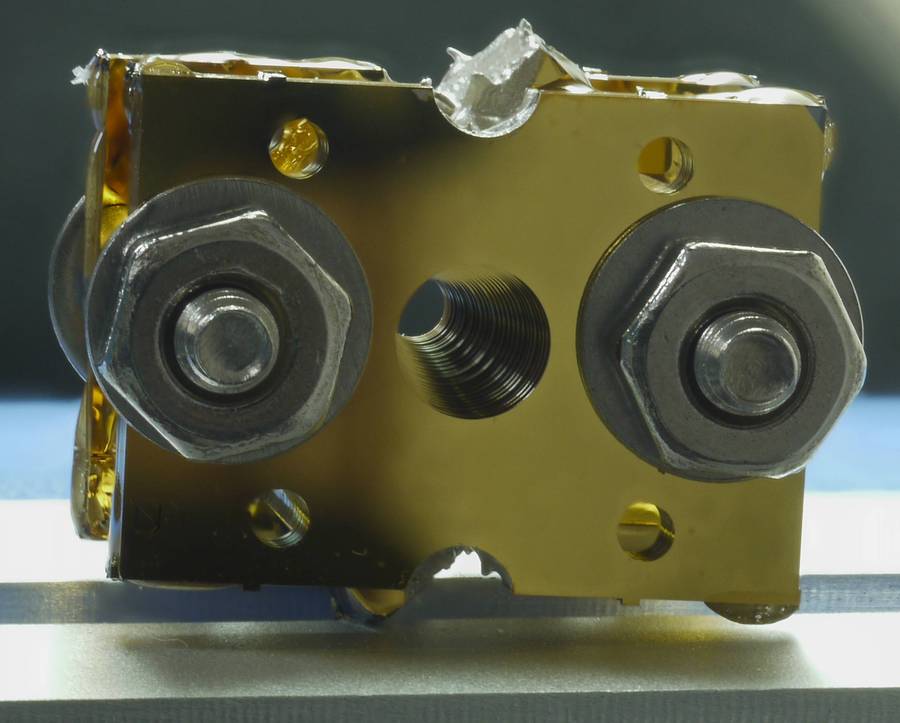}
\par\end{centering}

\caption{(left) Cross-section schematic of the prototype $150$ GHz corrugated
feed. The first six corrugation slots smoothly transform the fundamental
TE11 mode of the smooth-wall waveguide to the HE11 mode of the corrugated
section. A sine-squared profile is used in the flare section to the
aperture to achieve a more compact design than is possible with a
constant taper angle, with the added benefit of the location of the
phase center of the resulting beam that is coincident with the aperture
plane and is frequency-independent. (right) Photograph of a single
Si feed prior to electroplating. Features are (A) one of two clearance
holes filled with $4-40$ screws for holding together the platelets
during electroplating, (B) one of four holes for stainless alignment
dowels which mate with standard waveguides, and (C) is the horn aperture.
Corrugations are visible extending downward into the platelet stack
from the horn aperture. The translucent yellow is dried epoxy used
to adhere the platelets during electroplating. }

\label{fig:fh2schematicAndGlamorShot}
\end{figure}

The prototype corrugated horn design is driven by the needs of the
SPTpol and ACTpol receivers, both of which feature relatively fast
optics at the focal plane ($F\,\sim\,1.2$\textendash{}$1.3$). \cite{yoon2010spieTesPolarimetersForCMB,niemack2010ACTpol}
The input aperture diameter of the feed is determined by focal plane
sensitivity calculations, optimizing the tradeoff between beam spillover
efficiency and packing density. For the aforementioned receivers,
simulations indicate an optimum clear aperture diameter of $4.26\,\mbox{mm}$
(center-to-center packing distance of $5.26\,\mbox{mm}$ accounting
for the corrugation depths at the aperture), giving $34^{\circ}$
full width half maximum (FWHM) at nominal $150\,\mbox{GHz}$ operation.
The horn design follows standard practice. \cite{clarricoats1984corrugatedHorns,granet2005corrHornDesignPrimer}
It was optimized over the band from $122$~GHz to $170$ GHz, appropriate
for the $150$ GHz atmospheric window. The design was verified using
a modal-matching simulation package (Microwave Wizard by Mician, GmbH).\cite{nistComDisclaimer}
 See Fig.~\ref{fig:fh2schematicAndGlamorShot} for a schematic. 

The design was tested experimentally by fabrication and measurement
of a horn made from $60$ Si platelets each $250\,\mu\mbox{m}$ thick,
each corresponding to a ridge or a groove of a corrugation (see Figure~\ref{fig:fh2schematicAndGlamorShot}).
A metal seed layer (Ti:Au::$100\,\mbox{nm}$:$500\,\mbox{nm}$) was
deposited on the platelets mounted on an orbital platform canted at
$45^{\circ}$. Platelets were then stacked on stainless steel alignment
pins, clamped with a jig and adhered at the edges with an electrically
conductive epoxy. Platelet alignment accuracy by this approach was
$<\pm10\,\mu\mbox{m}$ layer-to-layer.\cite{britton2009ltd13siHornArray} 

Subsequent electroplating of $3\,\mu\mbox{m}$ thick Cu followed by
$3\,\mu\mbox{m}$ thick Au was performed to ensure a high-conductivity
finish and to fill gaps between platelets ensuring electrical continuity.
Note that the thickness of electrolytically deposited metals is systematically
thinner in hard-to-reach high-aspect crevasses such as the corrugations
near the center of the platelet array.\cite{britton2009ltd13siHornArray}
Cited thicknesses are as measured on flat, superficial surfaces. Electroplated
copper was selected as an underlying layer due to its gap-filling
properties. The Au plating thickness was selected to be well in excess
of the skin depth of bulk Au: $90$~nm at $100$~GHz and $273\,\mbox{K}$.\cite{kayelaby1995constantsWeb,jackson1999a}.

\section{FEED PERFORMANCE}

\label{sub:fhPerformance}

\begin{figure}
\begin{centering}
\includegraphics[width=5in]{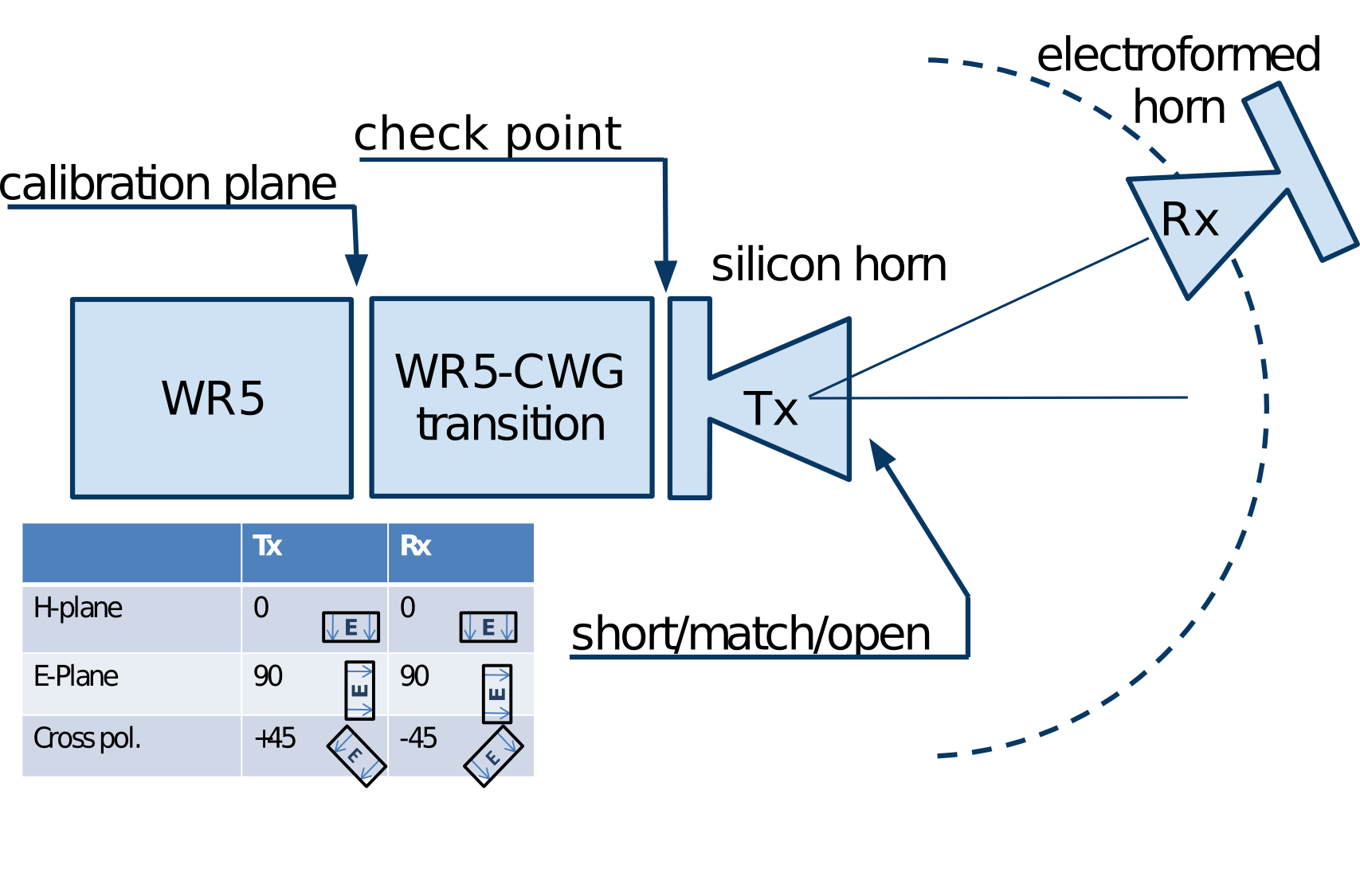}
\par\end{centering}

\caption{Schematic of feed test setup. It consists of two horns: a fixed transmitter
(Tx) and a receiver (Rx) that pivots about the transmitter's phase
center. The VNA S11 port is attached to the WR5 guide with a directional
coupler. A WR5-to-circular waveguide (CWG) transition interfaces with
the Tx horn. The VNA's S12 port is attached to the Rx horn. S11 was
calibrated at the calibration plane using a fixed short, sliding short
and sliding match. Since round calibration fixtures were unavailable,
horn measurements were compared with observation at the labeled check
point following the WR5-CWG transition. }

\label{fig:mmWaveSetup2}
\end{figure}

A vector network analyzer (VNA) configured as described in Fig.~\ref{fig:mmWaveSetup2}
was used to characterize our horn's return loss, insertion loss and
far-field radiation patterns. The apparatus consisted of two horns:
a fixed transmitter (Tx) and a receiver (Rx) that pivoted about the
transmitting horn's phase center. The Rx horn was a commercially manufactured
corrugated metal horn for G-band (WR5). Simulated cross polarization
is $<-30\,\mbox{dB}$ and side-lobe amplitude is $<-25$~dB at $150$,
$180$ and $220$ GHz. Return loss and insertion loss were measured
with the horn aperture terminated against a microwave absorber or
a shorting metal plate, respectively. To measure H-plane and E-plane
beam patterns we used a $0^{\circ}$ or $90^{\circ}$ twist preceding
the Tx and after the Rx horns. Cross polarization measurements used
$45^{\circ}$ twists before the Tx horn and after the Rx horn. To
confirm proper operation of the system we used a pair of identical
metal horns in a test beam pattern measurement sweep. Manufacturing
errors in the waveguide twists and our angle-sweeping setup limited
the polarization aligment accuracy to $\sim\pm1^{\circ}$, resulting
in the leakage of the co-polar beam patterns dominating the nominal
cross-polar beam pattern data at the level of $\sim-23$ dB; this
represents an upper limit on the cross-polar levels of the prototype
Si feed. 

\begin{figure}
\begin{centering}
\includegraphics[width=0.75\columnwidth]{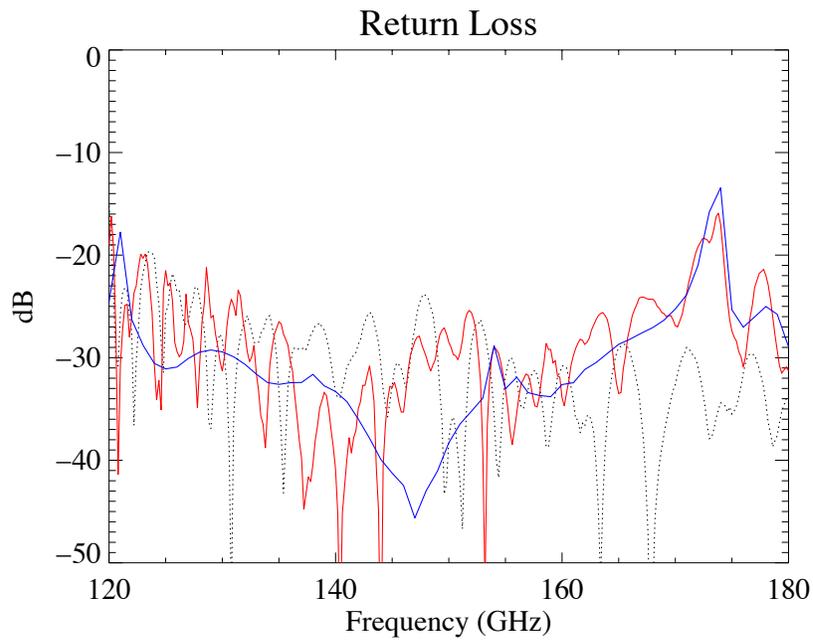}
\par\end{centering}

\caption{Simulated (solid, blue) and measured (solid, red) return loss. Also
shown (dotted) is the return loss of the rectangular-to-circular transition
necessary to mate to the Si feed. This transition is located beyond
the calibration point, and as such its intrinsic return loss contributes
to that of the Si feed's as shown above. See Fig.~\ref{fig:mmWaveSetup2}
for an explanation of the check point. The average return loss is
$<-20$~dB from $120$~GHz to $170$~GHz.}

\label{fig:fh2ReturnLoss}
\end{figure}

\begin{figure}
\begin{centering}
\includegraphics[width=0.75\columnwidth]{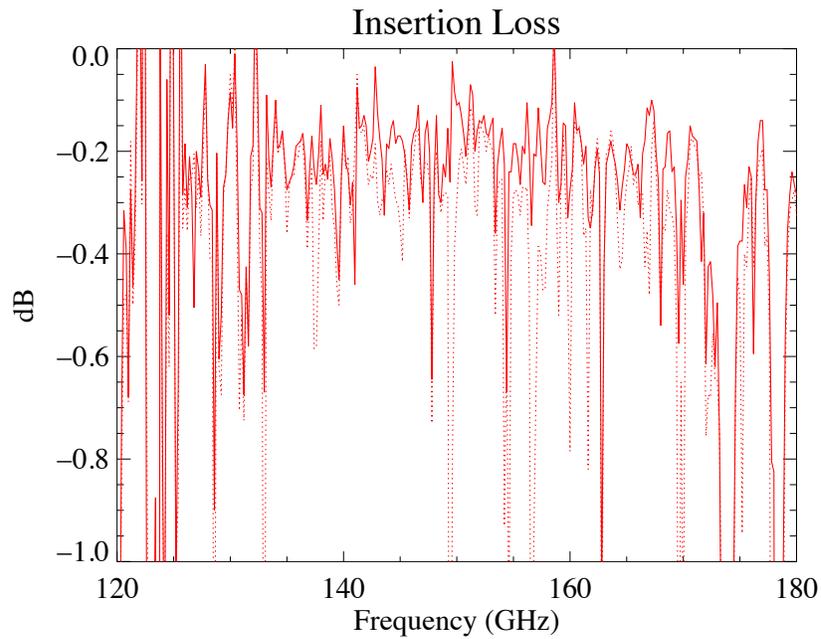}
\par\end{centering}

\caption{(solid trace) Implied insertion loss (IIL) is $(\mbox{S11}_{\mbox{horn}}^{2}-\mbox{S11}_{\mbox{CP}}^{2})/2$
where $\mbox{S11}_{\mbox{CP}}$ is measured with a short at the check
point (Fig.~\ref{fig:mmWaveSetup2}) and $\mbox{S11}_{\mbox{horn}}$
is measured with a short pressed against the output face of the horn.
This expression accounts for reflection from the WR5-CWG transition.
The average insertion loss is $\sim-0.2$~dB from $130$~GHz to
$170$~GHz. (dashed trace) IIL when the horn is intentionally misaligned
(lateral translation) with the circular waveguide (CWG) by $\sim0.1$
mm. }

\label{fig:fh2InsertionLoss}
\end{figure}

\begin{figure}
\begin{centering}
\includegraphics[width=0.9\columnwidth]{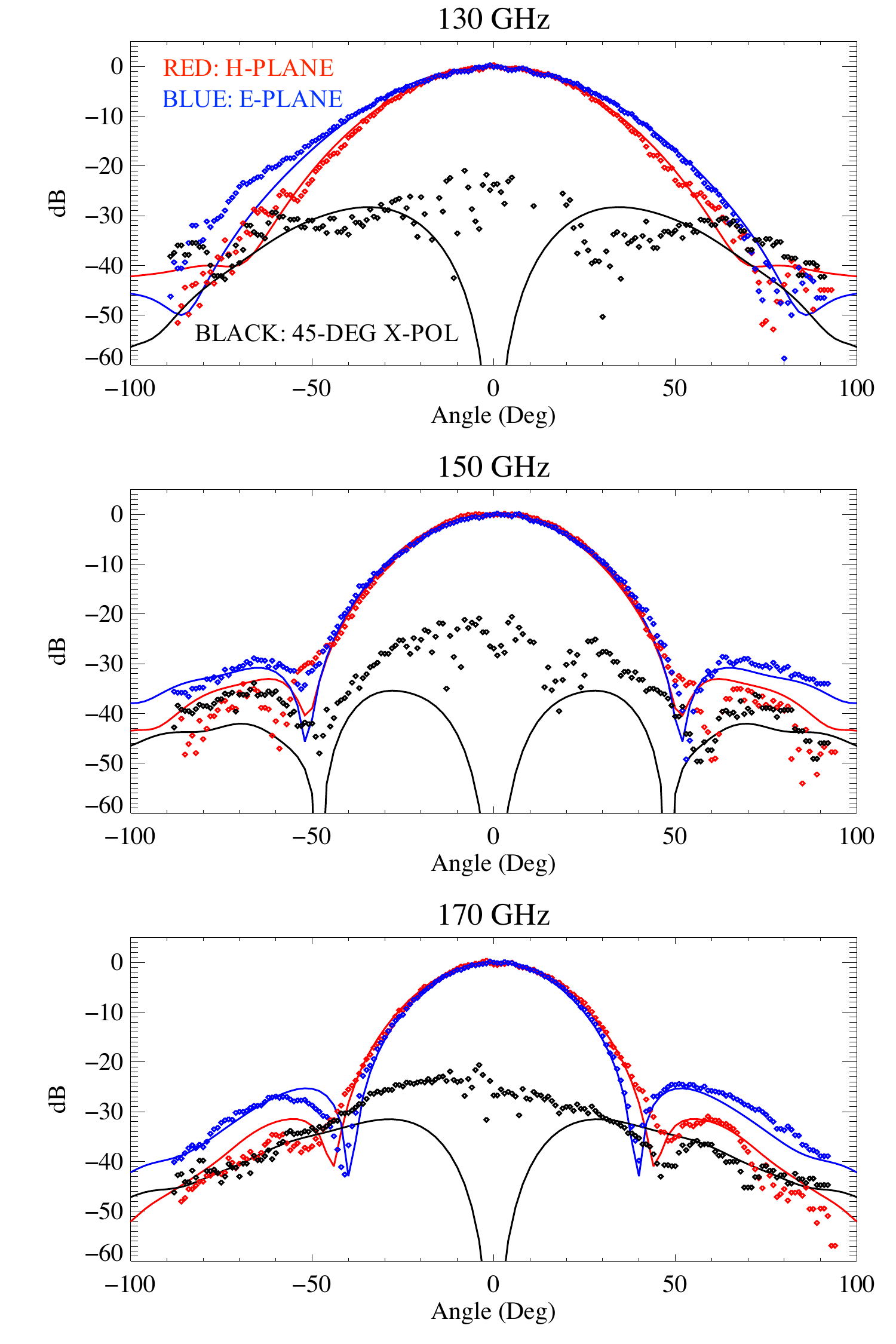}
\par\end{centering}

\caption{Plot of beam pattern and cross polarization at $130$, $150$ and
$170$~GHz. The solid traces are simulation while the points are
experiment. We conclude that from $130$~GHz to $170$~GHz the cross
polarization is $<-23$~dB, the side lobes are $<-25$~dB and beam
symmetry is good. }

\label{fig:fh2BeamPattern145GHz}
\end{figure}

Figures \ref{fig:fh2ReturnLoss}, \ref{fig:fh2InsertionLoss} and
\ref{fig:fh2BeamPattern145GHz} show the measured return loss, insertion
loss and far-field radiation pattern and cross polarization. 

After the microwave measurements the horn was sliced in cross section
on a dicing saw. This permitted inspection of the horn's interior
for micromachining, assembly and metalization defects. In particular
we checked that the electroplated metal conformally coats sharp corners
and bridges (fills) platelet-platelet gaps. Of $94$ inspected corners
none were found without adequate metalization. Of $90$ inspected
platelet-platlet junctions three were found with unsatisfactory coating
of which only one which was certainly gapped. 

\section{SILICON FEED ARRAY}

\begin{figure}
\begin{centering}
\includegraphics[width=0.9\columnwidth]{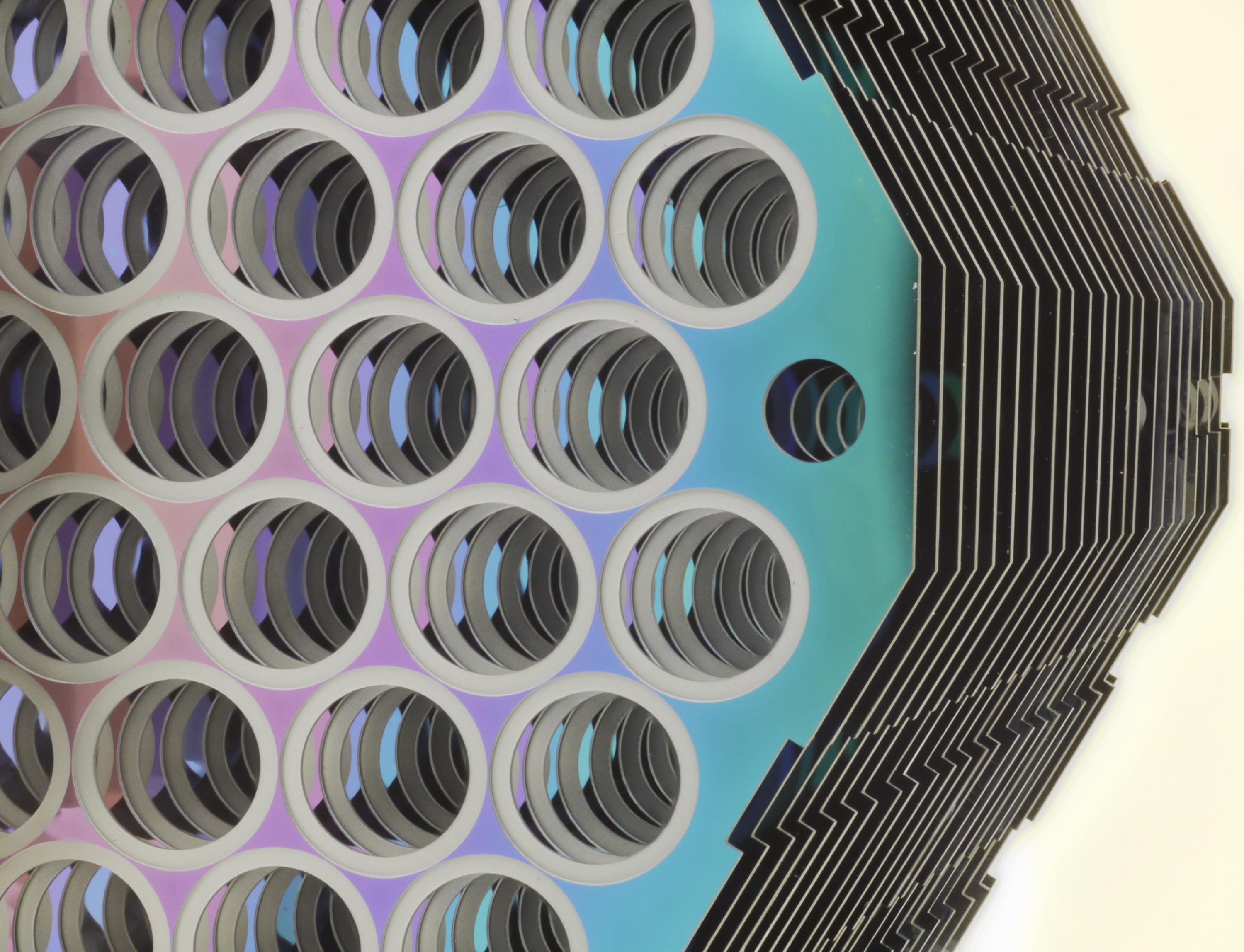}\caption{Perspective view of $23$ of the $33$ Si platelets that will be stacked
and metallized to form a $50\,\mbox{mm}$ diameter array with $84$~horns.
This planar array of corrugated platelet feeds will be integrated
with a matching arrays of NIST fabriacated OMT-coupled TES polarimeters
fabricated on Si and $\lambda/4$ Au-coated Si backshorts.}

\par\end{centering}

\label{fig:cmb6ExplodedView}
\end{figure}

In late $2009$ a prototype array structure with $73$ corrugated
waveguides was fabricated, temperature cycled (to $77\,\mbox{K}$)
and tested on the VNA. Testing methods and performance were similar
to the discrete Au-coated Si waveguides reported in $2009$.\cite{britton2009ltd13siHornArray}
Encouraged by the performance of this monolithic array of waveguides,
we are now fabricating a $50\,\mbox{mm}$ diameter array with $84$~horns
consisting of $33$~Si platelets (see Fig.~\ref{fig:cmb6ExplodedView}).

\begin{figure}
\begin{centering}
\includegraphics[width=0.8\columnwidth]{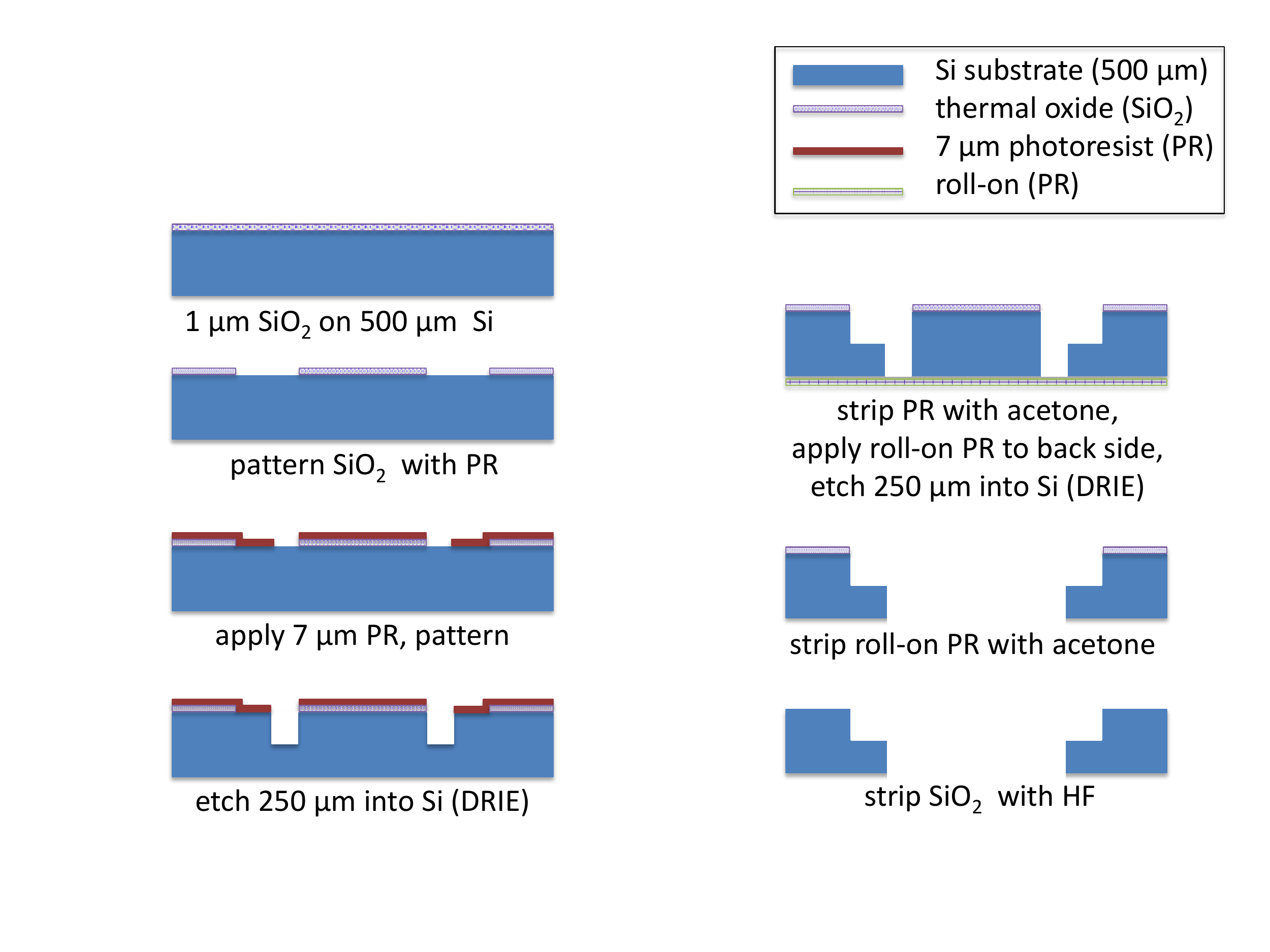}\caption{Figure illustrating the the process flow to fabricate platelets in
Si. For clarity only a single pixel is shown; in practice all pixels
are fabricated in parallel for each platelet. A pair of overlapping
etch masks and the additivity of the DRIE etch process were exploited.
The lower, wide-feature mask was defined in $\sim1\,\mu\mbox{m}$
thermal $\mbox{SiO}_{2}$ using photolithography and a HF wet etch.
The upper narrow-feature mask was defined in $7\,\mu\mbox{m}$ photoresist
and patterned using photolithography. At the outset of etching, both
masks were patterned and adhered to the wafer. After an initial $250\,\mu\mbox{m}$
etch using the narrow-feature mask, it was stripped off with acetone
leaving behind the wide-feature mask. Residual DRIE passivation material
was stripped with Dupont EKC-265 heated to $75$~C.\cite{nistComDisclaimer}
To protect the etcher's chuck, Dupont MX5020 roll-on photoresist was
applied to the back-side of the wafer.\cite{nistComDisclaimer} The
second $250\,\mu\mbox{m}$ deep etch penetrates the wafer. The remaining
photoresist with acetone and a Si plug at the core of each aperture
falls out. The $\mbox{SiO}_{2}$ is removed with a HF wet etch. }

\par\end{centering}

\label{fig:fabProcess}
\end{figure}

\begin{figure}
\begin{centering}
\includegraphics[width=0.49\columnwidth]{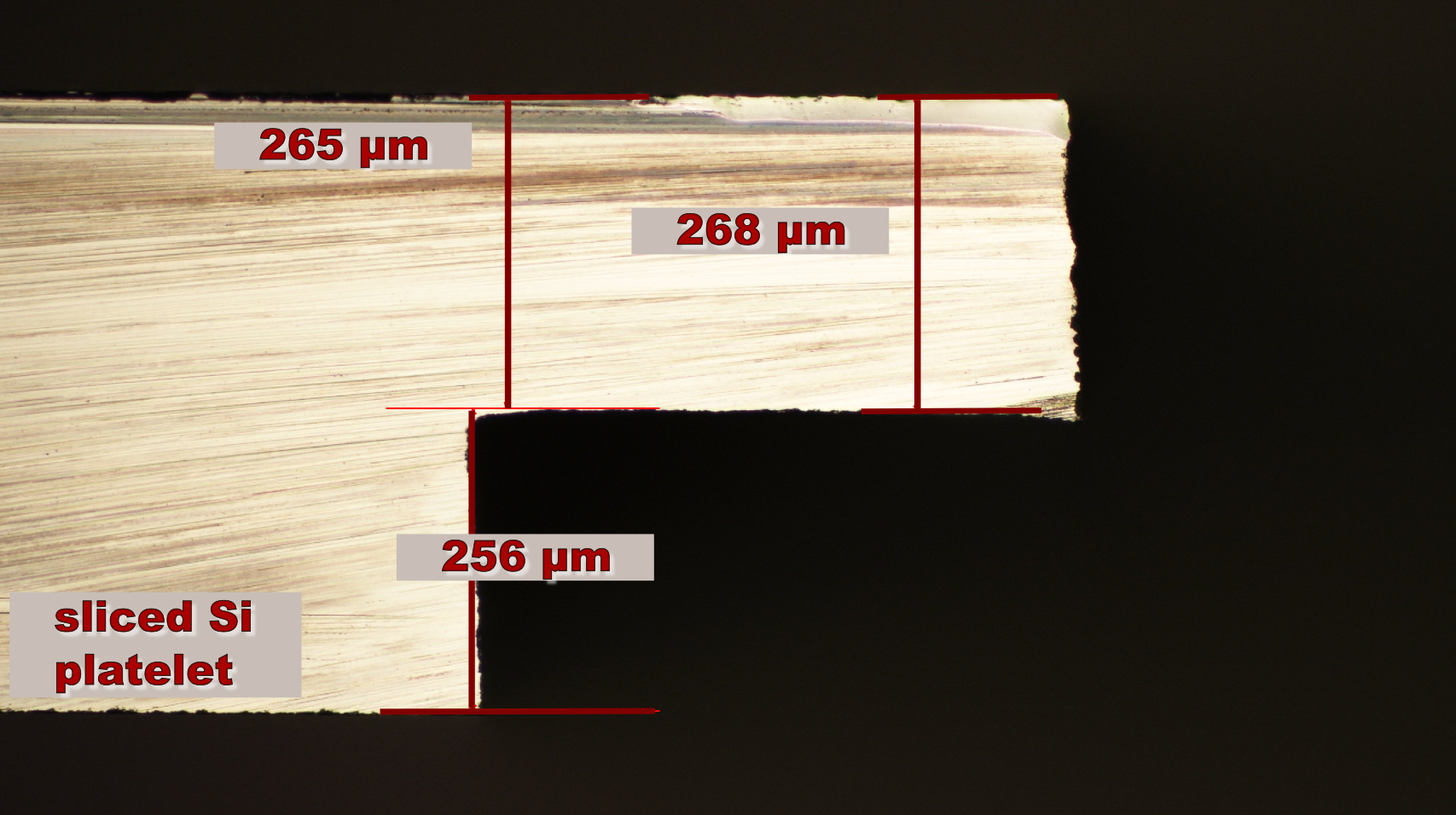} 
\par\end{centering}

\caption{(left) Photograph illustrating typical cross section of a corrugation.
Note that wafer thickness varies wafer to wafer: $525\pm25\,\mu\mbox{m}$.
The target etch depth was $250\,\mu\mbox{m}$ (measured from the bottom).
This wafer was not metallized. }

\label{fig:100mmWaferPics}
\end{figure}

For corrugated feeds it is necessary to use $>4$ corrugations per
$\lambda$.\cite{clarricoats1984corrugatedHorns} Higher frequencies
require thinner, more fragile wafers. However, thin, large diameter
wafers are prone to fracture; handling $150$~mm diameter wafers
thinner than $500\,\mu\mbox{m}$ is not desirable. To address this
problem a two-tiered etch protocol was developed to permit the use
of thicker wafers ($\sim\lambda/4$) each defining a full corrugation
period. To reduce wafer handling the etch proceeded from a single
side in two stages. See Figure~\ref{fig:fabProcess} for a description
of the Si platelet fabrication process and Figure~\ref{fig:100mmWaferPics}
for images of Si platelets.

Typical wafer parameters for the Si used in the platelets are $0.001-0.005\,\Omega\cdot\mbox{cm}$
bulk resistivity (B doped, p-type), <100> crystal orientation (denoted
by wafer flat), double-side polished ($<0.5$~nm rms), $525\pm25\,\mu\mbox{m}$
thickness and $<20\,\mu\mbox{m}$ bow. 

Individual platelets require metalization prior to stacking to facilitate
subsequent electrolytic plating. Metal deposition by evaporation proved
very robust in tests on individual feeds (Ti:Au::$100\,\mbox{nm}$:$500\,\mbox{nm}$)
and will be used for monolithic arrays as well. As with the individual
platelet feeds, platelet-platelet registration will be performed using
(removable) stainless dowels and clearance holes micromachined in
the Si. We are exploring use of epoxy and spring loaded clamps to
ensure that the the gap between platelets is minimal during electroplating
(Cu:Au::$3\,\mu\mbox{m}$:$3\,\mu\mbox{m}$).

\section{CONCLUSION}

\label{sec:conclusion}

We fabricated and tested Au-coated corrugated Si platelet waveguides,
waveguide arrays and horns. These devices exhibited performance comparable
to conventional all-metal devices. Our current work focuses on extending
these techniques to fabrication of monolithic waveguide arrays with
$600+$ individual pixels on $150$~mm wafers. 

The flexibility of Si micromachining makes possible a wide varity
of microwave structures. For example, the 2-tiered etch recipe permits
overlap of adjacent horns' corrugations at the horn apertures for
greater packing density. A 3-tiered etch makes posible {}``ring-loaded''
corrugated horns with greatly increased spectral bandwidth for potential
multichroic detectors.\cite{mcmahon2009ringLoadedConversation,takeichi1971ringLoadedWaveguide}

\section*{APPENDIX A. SILICON DEEP ETCH}

\label{app:SiDRIE}

Plasma etching of silicon permits high aspect ratio features with
good repeatability and excellent mask selectivity. The NIST etcher
utilizes a variant of the BOSCH etch process optimized for silicon
deep etch ($>50\,\mu\mbox{m}$).\cite{larmer1992a,mcauley2001a} This
process forms nearly vertical sidewalls in silicon by interleaving
etch and surface passivation steps. The etch step is a chemically
active RF plasma ($\mbox{SF}_{6}$ and $\mbox{O}_{2}$) inductively
coupled to the wafer surface. The charged component of the plasma
is accelerated normal to the surface, enhancing its etch rate in the
vertical direction (tunable, $1-30\,\mu\mbox{m/min}$). The passivation
step ($\mbox{C}_{4}\mbox{F}_{8}$) coats all exposed surfaces including
sidewalls with a fluorocarbon polymer. Etch and passivation cycles
(typically $12$ and $8$ seconds respectively) are balanced so that
sidewalls are protected from over/under etching as material is removed
through the full wafer thickness. The cycled etching causes the sidewalls
to have a microscopic scalloped appearance with an amplitude of $100-500\,\mbox{nm}$
and a period of $200-1000\,\mbox{nm}$.\cite{mcauley2001a}

The primary etch mechanism for the $\mbox{SF}_{6}$/$\mbox{O}_{2}$
chemistry is due to chemical reactions between neutral atomic Fluorine
formed in the plasma and silicon at the wafer surface. Oxygen acts
to degrade the Fluorocarbon polymers ($\mbox{CF}_{x}$) to form gaseous
products. While other etch chemistries are possible, this is the most
common owing to the safety of the reactants and their selectivity
of silicon over commonly used etch masks (photoresist and silicon
oxide).\cite{flamm1990a,madou2002a} The gaseous products are pumped
away and do not redeposit on the process wafer. Structures with multiple
tiers per wafer are possible through the use of overlapping etch masks\cite{britton2009aplSiTraps}
or, at lower resolution, with roll-on photoresist. 

DRIE permits fabrication of high-apect features in Si ($\sim50:1$
aspect ratio). Lateral resolution is limited by the greater of photolithography
resolution and etch aspect ratio. For example, in $250\,\mu\mbox{m}$
thick Si the NIST etcher's lateral resolution is $\pm5\,\mu\mbox{m}$
through the full wafer bulk even though our photolithography resolution
is $\sim3\,\mu\mbox{m}$. Radial variation in etch rate is ultimately
determined by aspects of the DRIE etch tool such as plasma uniformity;
we observe a radial (center to edge) variation of $<3\%$. Tools for
applying this process to wafers $150\,\mbox{mm}$ in diameter are
currently available in the NIST microfabrication facility. High resolution
wafer-scale Si etch tools are ubiquitous in Si MEMs foundries.

\section*{ACKNOWLEDGMENTS}

We thank the members of the ACT and SPT collaborations for their support
and interest in this research. We thank John J. Jost and Dave Walker
for suggestions on the manuscript. M.~D.~Niemack and K.~W.~Yoon
acknowledge support from the National Research Council. This work
was supported by NIST through the Innovations in Measurement Science
program. This proceeding is a contribution of NIST and is not subject
to U.S. copyright. 

\clearpage

\bibliographystyle{spiebib}
\bibliography{/Users/britton/Dropbox/spie2010/LinkTojabRef/joe_b_v3}

\begin{thebibliography}{10}
\bibitem{nistComDisclaimer}
``{Products or companies named here are cited only in the interest of complete
  technical description, and neither constitute nor imply endorsement by {NIST}
  or by the {US} government. Other products may be found to serve just as
  well}.''
\end{thebibliography}

\end{document}